\documentclass[prl,aps,twocolumn,floatfix,showpacs]{revtex4}
\usepackage{graphics}
\begin{document}
\title{Trap-imbalanced fermion mixtures}
\author{M. Iskin and C. J. Williams}
\affiliation{Joint Quantum Institute, National Institute of Standards and Technology and University of Maryland, Gaithersburg, Maryland 20899-8423, USA.}
\date{\today}

\begin{abstract}
We analyze the ground state phases of two-component ($\sigma  \equiv \lbrace \uparrow, \downarrow \rbrace$) 
population- and mass-balanced ($N_\uparrow = N_\downarrow$ and $m_\uparrow = m_\downarrow$) 
but trap-imbalanced ($\omega_\uparrow \ne \omega_\downarrow$)
fermion mixtures as a function of interaction strength from the weak
attraction Bardeen-Cooper-Schrieffer (BCS) to the strong attraction 
Bose-Einstein condensation (BEC) limit.
In the BCS limit, we find that the unpolarized superfluid (UPS) fermions exist
away from the central core of the trapping potentials, and
are surrounded by partially polarized normal (P$\sigma$PN) fermions.
As the interactions increase towards unitarity, we find that the central P$\sigma$PN core
first transitions to a UPS, and then expands towards the edges until the 
entire mixture becomes a UPS in the BEC limit.

\end{abstract}
\maketitle

Ultracold atomic physics experiments with two-component fermion mixtures have enabled the study 
of novel superfluid and insulating phases which have not been possible in other systems.
For instance, the tuning of attractive fermion-fermion interactions have permitted the ground 
state of the system to evolve from a weak fermion attraction Bardeen-Cooper-Schrieffer (BCS) 
limit of loosely bound and largely overlapping Cooper pairs to a strong fermion attraction 
limit of tightly bound and small bosonic molecules which undergo Bose-Einstein condensation
(BEC)~\cite{chin, regal2, bourdel1, partridge1, kinast2, zwierlein3}.
For mass- and population-balanced mixtures, in agreement with the early theoretical 
predictions~\cite{leggett}, these experiments have shown that the BCS-BEC evolution is just a crossover.

Recently, the ground state phase diagram of mass-balanced but population-imbalanced
fermion mixtures have been theoretically analyzed showing that the BCS-BEC 
evolution is not a crossover but quantum phase transitions occur between normal and superfluid 
phases~\cite{pao, sheehy}. In addition, phase separation between 
superfluid (paired) and normal (excess) fermions has been shown.
Motivated by these predictions, there have been several experiments with
mass-balanced but population-imbalanced fermion mixtures~\cite{mit, rice}, leading to an 
intensive theoretical activity~\cite{torma, pieri, yi, silva, haque, lobo, liu-mixture, mizushima}. 
Since exotic superfluid phases (i.e., Fulde-Ferrell and Larkin-Ovchinnikov (FFLO)~\cite{FF, LO}) 
can be potentially realized, imbalanced fermion mixtures are currently interest to 
many communities ranging from atomic and molecular to condensed- and nuclear-matter physics.
For instance, the ground state phase diagram of mass- and population-imbalanced fermion mixtures 
have been recently analyzed showing quantum and topological 
phase transitions~\cite{iskin-mixture, pao-mixture, duan-mixture, parish, paananen, pao-unequal}.

\begin{figure} [htb]
\centerline{\scalebox{0.3}{\includegraphics{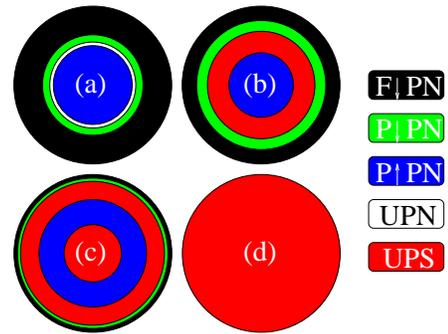}}}
\caption{\label{fig:trap.shells} (Color online)
Schematic diagrams showing shell-structures of two-component 
($\sigma  \equiv \lbrace \uparrow, \downarrow \rbrace$)
population- and mass-balanced ($N_\uparrow = N_\downarrow$ and $m_\uparrow = m_\downarrow$) but 
trap-imbalanced ($\omega_\uparrow > \omega_\downarrow$) fermion mixtures for 
(a) non-interacting, and
(b) weak,
(c) intermediate and
(d) strong attraction regimes.
Here, $\omega_\sigma$ is the trapping frequency of $\sigma$-fermions.
The colored regions correspond to
unpolarized superfluid (UPS, red),
unpolarized normal (UPN, white),
partially $\uparrow$-polarized normal (P$\uparrow$PN, blue),
partially $\downarrow$-polarized normal (P$\downarrow$PN, green), and
fully $\downarrow$-polarized normal (F$\downarrow$PN, black) phases.
}
\end{figure}

In this manuscript, we analyze the ground state phases of
two-component ($\sigma  \equiv \lbrace \uparrow, \downarrow \rbrace$) population- 
and mass-balanced ($N_\uparrow = N_\downarrow$ and $m_\uparrow = m_\downarrow$) 
but trap-imbalanced ($\omega_\uparrow \ne \omega_\downarrow$) fermion mixtures
as a function of interaction strength,
where $\omega_\sigma$ is the trapping frequency of $\sigma$-component.
The ground state involves very rich shell-structures consisting of unpolarized superfluid (UPS) 
and unpolarized normal (UPN) as well as partially $\sigma$-polarized normal (P$\sigma$PN)
and fully $\sigma$-polarized normal (F$\sigma$PN) fermions.
Our results are schematically shown in Fig.~\ref{fig:trap.shells}, and are as follows.
In the BCS limit shown in Fig.~\ref{fig:trap.shells}(b), we find that the UPS 
fermions exist only away from the central core of the trapping potentials, and are surrounded
by P$\sigma$PN fermions. As the interactions increase towards unitarity, we find that
the central P$\sigma$PN core first transitions to a UPS as shown in Fig.~\ref{fig:trap.shells}(c), 
and then expands towards the edges until the entire mixture becomes a UPS in the BEC limit 
as shown in Fig.~\ref{fig:trap.shells}(d).

{\it Pairing Hamiltonian}:
To obtain these results, we start with the Hamiltonian density (in units of $\hbar = k_B = 1$),
\begin{equation}
\label{eqn:hamiltonian}
H(\mathbf{r}) = \sum_{\sigma} \psi_{\sigma}^\dagger (\mathbf{r}) K_\sigma(\mathbf{r}) \psi_{\sigma}(\mathbf{r})
- g \Psi_{\uparrow,\downarrow}^\dagger (\mathbf{r}) \Psi_{\uparrow,\downarrow}(\mathbf{r}),
\end{equation}
where $\psi_{\sigma}^\dagger (\mathbf{r})$ creates a pseudo-spin-$\sigma$ 
fermion at position $\mathbf{r}$, and
$
\Psi_{\uparrow,\downarrow}^\dagger (\mathbf{r}) = 
\psi_{\uparrow}^\dagger (\mathbf{r}) \psi_{\downarrow}^\dagger (\mathbf{r})
$
is the pair creation operator.
In Eq.~(\ref{eqn:hamiltonian}), $g > 0$ is the strengh of the attractive 
fermion-fermion interactions, and we defined
$
K_\sigma(\mathbf{r}) = - \nabla^2/(2m_{\sigma})  - \mu_\sigma(\mathbf{r})
$
where
$
\mu_\sigma(\mathbf{r}) = \mu_\sigma - V_\sigma(\mathbf{\mathbf{r}})
$
is the local chemical potential.
The global chemical potentials $\mu_{\sigma}$ fixes the density $n_{\sigma} = N_\sigma/V$ 
of each type of fermion independently, where $N_\sigma$ is the number of 
$\sigma$-fermions and $V$ is the volume.
The term
$
V_\sigma(\mathbf{r}) = m_\sigma (\omega_{\sigma,x}^2 x^2 + \omega_{\sigma,y}^2 y^2 + \omega_{\sigma,z}^2 z^2)/2
$
corresponds to the trapping potential, which is assumed to be harmonic in space.

In the momentum space, within the local-density (LD) approximation, 
the local mean-field (MF) Hamiltonian can be written as
$
H_\mathbf{Q}(\mathbf{r}) = \sum_{\mathbf{k},\sigma} \xi_{\mathbf{k},\sigma}(\mathbf{r}) 
a_{\mathbf{k}, \sigma}^\dagger a_{\mathbf{k}, \sigma}
- \Delta_\mathbf{Q}(\mathbf{r}) \sum_{\mathbf{k}}
(a_{\mathbf{k}+\mathbf{Q}/2,\uparrow}^\dagger 
a_{-\mathbf{k}+\mathbf{Q}/2,\downarrow}^\dagger + h.c.) 
+ \Delta_\mathbf{Q}^2(\mathbf{r})/g,
$
where $\mathbf{Q}$ is the center-of-mass momentum of individual Cooper pairs, and
$
\xi_{\mathbf{k},\sigma}(\mathbf{r})= \epsilon_{\mathbf{k},\sigma} - \mu_\sigma(\mathbf{r})
$ 
with 
$
\epsilon_{\mathbf{k},\sigma} = |\mathbf{k}|^2/(2m_\sigma).
$
Here, $\Delta_\mathbf{Q}(\mathbf{r})$ is the local MF order parameter 
which is assumed to be real without loss of generality, and defined by
$
\Delta_\mathbf{Q}(\mathbf{r}) = g \sum_{\mathbf{k}} 
\langle 
a_{-\mathbf{k}+\mathbf{Q}/2,\downarrow}
a_{\mathbf{k}+\mathbf{Q}/2,\uparrow}
\rangle,
$
where $\langle . \rangle$ implies a thermal average.

{\it Self-Consistency Equations}:
The local MF Hamiltonian can now be solved by using standard techniques~\cite{pao, iskin-mixture, liu-mixture}.
The order parameter $\Delta_{\mathbf{Q}}(\mathbf{r})$ is determined by
\begin{equation}
\frac{M V}{4\pi a_F} = \sum_{\mathbf{k}} \bigg\lbrace \frac{1}{2\epsilon_{\mathbf{k}}}
- \frac{1 - f[E_{\mathbf{k},\mathbf{Q},\uparrow}(\mathbf{r})] - f[E_{\mathbf{k},\mathbf{Q},\downarrow}(\mathbf{r})]} 
{2E_{\mathbf{k},\mathbf{Q}}(\mathbf{r})} \bigg\rbrace,
\label{eqn:op}
\end{equation}
where
$
\epsilon_{\mathbf{k}} = (\epsilon_{\mathbf{k},\uparrow}+ \epsilon_{\mathbf{k},\downarrow})/2
$
is the average kinetic energy,
$
f(x) = 1/[\exp(x/T) + 1]
$
is the Fermi function, and
$
E_{\mathbf{k},\mathbf{Q}}(\mathbf{r}) = [\xi_{\mathbf{k},\mathbf{Q}}^2(\mathbf{r}) + \Delta_\mathbf{Q}^2(\mathbf{r})]^{1/2}
$
with
$
\xi_{\mathbf{k},\mathbf{Q}}(\mathbf{r}) = [\xi_{\mathbf{k}+\mathbf{Q}/2,\uparrow}(\mathbf{r}) + \xi_{-\mathbf{k}+\mathbf{Q}/2,\downarrow}(\mathbf{r})]/2.
$
Here,
$
E_{\mathbf{k},\mathbf{Q},\sigma}(\mathbf{r}) = E_{\mathbf{k},\mathbf{Q}}(\mathbf{r}) 
+ s_\sigma [\xi_{\mathbf{k}+\mathbf{Q}/2,\uparrow}(\mathbf{r})-\xi_{-\mathbf{k}+\mathbf{Q}/2,\downarrow}(\mathbf{r})]/2
$
is the quasi-particle energy when $s_\uparrow = 1$ or
the negative of the quasi-hole energy when $s_\downarrow = -1$.
Notice that, we elliminate $g$ in favor of the fermion-fermion 
scattering length $a_F$ via the usual regularization
$
1/g = - MV/(4\pi a_F) + \sum_{\mathbf{k}} 1/(2\epsilon_{\mathbf{k}}),
$
where $M = 2m_\uparrow m_\downarrow/(m_\uparrow + m_\downarrow)$ is twice the 
reduced mass of $\uparrow$- and $\downarrow$-fermions.
Eq.~(\ref{eqn:op}) has to be solved self-consistently with the number equations
$
N_\sigma = \int d\mathbf{r} n_\sigma(\mathbf{r}),
$
where 
\begin{eqnarray}
n_{\sigma}(r) &=& \frac{1}{V}\sum_{\mathbf{k}}
\big\lbrace u_{\mathbf{k},\mathbf{Q}}^2(\mathbf{r}) f[E_{\mathbf{k},\mathbf{Q},\sigma}(r)] \nonumber \\
&+& v_{\mathbf{k},\mathbf{Q}}^2(\mathbf{r}) f[-E_{\mathbf{k},\mathbf{Q},-\sigma}(\mathbf{r})] \big\rbrace,
\label{eqn:number}
\end{eqnarray}
is the local density of $\sigma$-fermions. Here, 
$
u_{\mathbf{k}, \mathbf{Q}}^2(\mathbf{r}) = 
[1 + \xi_{\mathbf{k}, \mathbf{Q}}(\mathbf{r})/E_{\mathbf{k}, \mathbf{Q}}(\mathbf{r})]/2
$
and
$
v_{\mathbf{k}, \mathbf{Q}}^2(\mathbf{r}) = 
[1 - \xi_{\mathbf{k}, \mathbf{Q}}(\mathbf{r})/E_{\mathbf{k}, \mathbf{Q}}(\mathbf{r})]/2.
$
Since the pseudo-spin symmetry is broken in population-, mass- and/or trap-imbalanced 
Fermi gases, one needs to solve all three equations self-consistently for all $\mathbf{Q}$,
and determine the value of $\mathbf{Q}$ which minimizes the free energy~\cite{FF, LO, sheehy}.

Having established the theoretical formalism, next we analyze the ground state
phases of trap-imbalanced fermion mixtures as a function of $a_F$. 
For this purpose, we first discuss the numerical results and then provide 
analytical insight into the problem.

\begin{figure} [htb]
\centerline{\scalebox{0.38}{\hskip 0mm \includegraphics{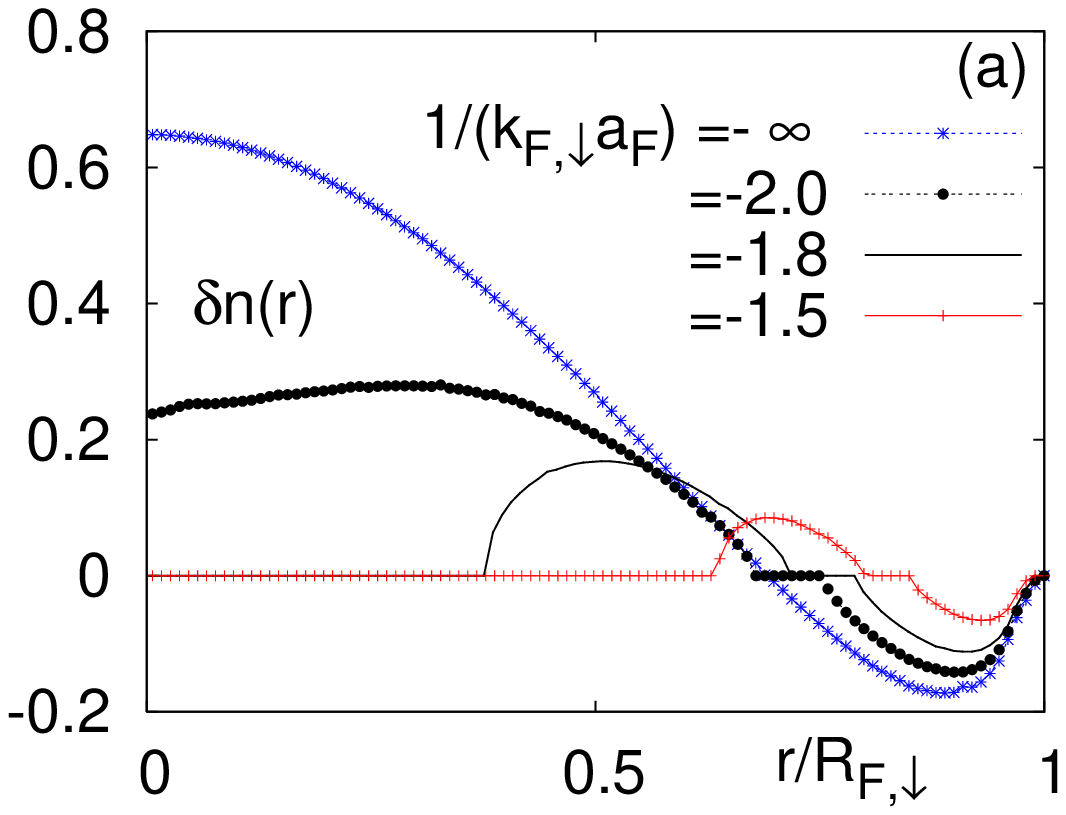} \hskip -20mm \includegraphics{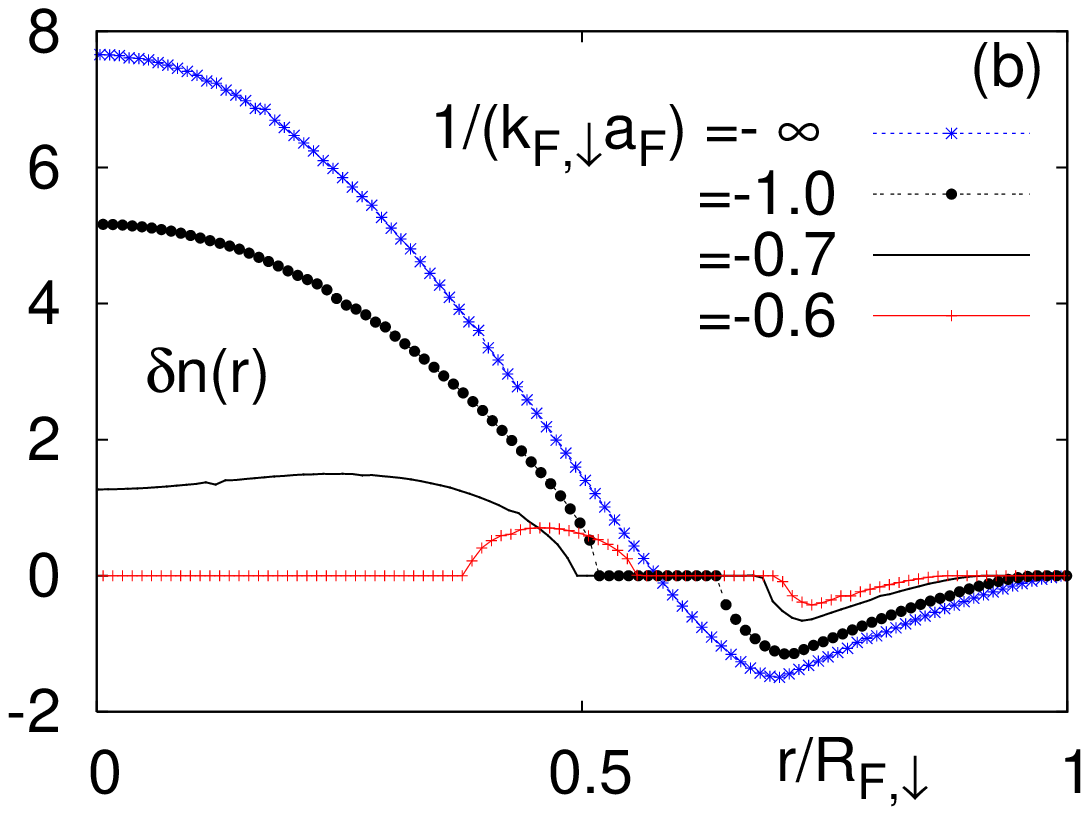}}}
\caption{\label{fig:trap.di} (Color online)
Local polarization $\delta n(r) = n_\uparrow (r) - n_\downarrow (r)$ 
[in units of $k_{F,\downarrow}^3/(2\pi)^3$] versus radius $r$ (in units of $R_{F,\downarrow}$) 
is shown for population- and mass-balanced ($N_\uparrow = N_\downarrow$ and $m_\uparrow = m_\downarrow$) 
but trap-imbalanced fermion mixtures, where the trapping frequencies are
(a) $\omega_\uparrow = 1.1 \omega_\downarrow$ and
(b) $\omega_\uparrow = 2 \omega_\downarrow$.
While the unpolarized regions are UPS, the polarized regions include
both FFLO superfluid, and P$\sigma$PN and F$\sigma$PN phases.
}
\end{figure}

{\it Numerical Calculations}:
In this manuscript, we assume that the trapping potentials are isotropic in space such that
$
V_\sigma(\mathbf{r}) = m_\sigma \omega_\sigma^2 r^2/2
$
where $r = |\mathbf{r}|$.
In addition, we do not explicitly consider the FFLO-like ($Q \ne 0$) 
superfluid phase~\cite{FF, LO}, and limit the numerical calculations to the $Q = 0$ phase.
However, this phase may be present in the weakly-attracting imbalanced 
fermion mixtures, but only in a narrow parameter space, as discussed below.
We also set $N_\uparrow = N_\downarrow$ and $m_\uparrow = m_\downarrow$, and
consider two cases (a) $\omega_\uparrow = 1.1 \omega_\downarrow$ and (b) $\omega_\uparrow = 2 \omega_\downarrow$.
The numerical calculation involves self-consistent solutions of Eqs.~(\ref{eqn:op}) and~(\ref{eqn:number})
for $\Delta_0 (r)$, $n_\sigma (r)$ and $\mu_\sigma$.
For instance, in Fig.~\ref{fig:trap.di}, we show the local polarization
$\delta n(r) = n_\uparrow (r) - n_\downarrow (r)$ as a function of $a_F$, 
characterizing the non-interacting, and weak, intermediate and strong attraction regimes.

We numerically find that the ground state involves very 
rich shell-structures consisting of UPS and UPN as well as P$\sigma$PN and F$\sigma$PN phases,
depending on the particular value of $a_F$ as shown in Fig.~\ref{fig:trap.shells}. 
To understand these shell-structures, next we analyze the non-interacting, and 
weakly- and strongly-attracting limits, which are analytically tractable.

{\it Non-Interacting Fermion Mixtures}:
To understand the interacting trap-imbalanced fermion mixtures, 
it is useful to analyze first the non-interacting case when $g = 0$ or $a_F \to 0^-$. 
In this limit, the mixture is in normal phase such that the superfluid order parameter 
vanishes at all space $\Delta_{Q}(r) = 0$, and that the global chemical 
potentials are identical to the global Fermi energies $\mu_\sigma = \epsilon_{F,\sigma}$
at zero temperature ($T = 0$).
Thus, Eq.~(\ref{eqn:number}) reduces to
$
n_\sigma(r) = (1/V) \sum_{\mathbf{k}} f[\xi_{\mathbf{k},\sigma}(r)],
$
and at $T = 0$ is given by 
$
n_\sigma (r) = (1/V) \sum_{k < k_{F,\sigma} (r)} 1,
$
where $k_{F,\sigma}(r)$ is the local Fermi momentum
defined by
$
\epsilon_{F,\sigma} = k_{F,\sigma}^2/(2m_\sigma) 
= k_{F,\sigma}^2(r)/(2m_\sigma) + V_\sigma(r).
$
This leads to
$
n_\sigma(r) = m_\sigma^3 \omega_\sigma^3 (R_{F,\sigma}^2 - r^2)^{3/2}/(6\pi^2),
$
where $R_{F,\sigma}$ is the Thomas-Fermi radius of $\sigma$-fermions defined by
$
\epsilon_{F,\sigma} = m_\sigma \omega_\sigma^2 R_{F,\sigma}^2/2,
$
such that $k_{F,\sigma} = m_\sigma \omega_\sigma R_{F,\sigma}$ is the global Fermi momentum.
Then, the number of $\sigma$-fermions is found by integrating 
$n_\sigma(r)$ over $\mathbf{r}$ where $r \le R_{F,\sigma}$, leading to
$
N_\sigma = k_{F,\sigma}^3 R_{F,\sigma}^3/48. 
$
Setting $N_\uparrow = N_\downarrow$ gives
$
R_{F,\uparrow}/R_{F,\downarrow} = [m_\downarrow \omega_{\downarrow}/(m_{\uparrow} \omega_{\uparrow})]^{1/2},
$
and therefore, a {\it trap-imbalanced fermion mixture} can be realized when the condition
$
m_\uparrow \omega_\uparrow \ne m_\downarrow \omega_\downarrow
$
is satisfied.

\begin{figure} [htb]
\centerline{\scalebox{0.3}{\includegraphics{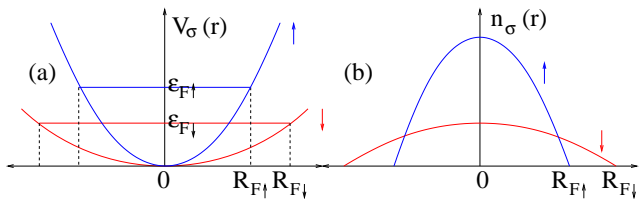}}}
\caption{\label{fig:trap.density} (Color online)
Schematic (a) trap ($V_\sigma)$ and (b) density ($n_\sigma$) profiles
are shown for non-interacting [$1/(k_{F,\downarrow} a_F) = - \infty$] population- and mass-balanced 
($N_\uparrow = N_\downarrow$ and $m_\uparrow = m_\downarrow$) but trap-imbalanced 
($\omega_\uparrow > \omega_\downarrow$) fermion mixtures.
}
\end{figure}

When $m_\uparrow \omega_\uparrow > m_\downarrow \omega_\downarrow$, since $R_{F,\uparrow} > R_{F,\downarrow}$, 
there are more $\uparrow$-fermions near the center of the 
trap while $\downarrow$-fermions are in excess near the edges as shown in Fig.~\ref{fig:trap.density}. 
In addition, the local density of $\uparrow$- and $\downarrow$-fermions
are equal only at radius 
$
r_c = R_{F,\downarrow} [m_\downarrow \omega_\downarrow (\omega_\uparrow-\omega_\downarrow)/(m_\uparrow \omega_\uparrow^2 - m_\downarrow \omega_\downarrow^2)]^{1/2},
$
satisfying $\delta n (r_c) = 0$.
Therefore, in this case, the ground state corresponds to a 
P$\uparrow$PN for $0 \le r < r_c$, to a UPN for $r = r_c$, to a
P$\downarrow$PN for $r_c < r \le R_{F,\uparrow}$, and to a
F$\downarrow$PN for $R_{F,\uparrow} < r \le R_{F,\downarrow}$.
For instance, when $\omega_\uparrow = 2 \omega_\downarrow$, we find $r_c \approx 0.58 R_{F,\downarrow}$ 
and $R_{F,\uparrow} \approx 0.71 R_{F,\downarrow}$, and these three phases can be seen 
in Fig.~\ref{fig:trap.di}(b) when $1/(k_{F,\downarrow} a_F) = -\infty$.
The shell-structure of this case is schematically shown in Fig.~\ref{fig:trap.shells}(a).

Having finite and attractive fermion-fermion interactions changes this simple 
non-interacting picture dramatically, which is discussed next.

{\it Weakly-Attracting Fermion Mixtures}:
When $g > 0$ or $a_F \lesssim 0$, the normal mixture may become unstable against 
formation of Cooper pairs at some regions of the trap such that $\Delta_{Q} (r) \ne 0$.
According to the BCS theory of superconductivity, this is always the case for chemical 
potentially balanced mixtures,
$
\delta \mu (r) = [\mu_\uparrow (r) - \mu_\downarrow (r)]/2 = 0
$
for all $r$, no matter how weak the $g$ is. 
This suggests that, for an arbitrarilly small $g$, trap-imbalanced fermion mixtures
first become unstable against superfluidity at radius $r_c$, where $\delta n (r_c) = 0$ 
and $\delta \mu (r_c) = 0$. 

In the weakly-attracting limit  when $g \ll \{\epsilon_{F,\uparrow}, \epsilon_{F,\downarrow}\}$,
the local order parameter is obtained from Eq.~(\ref{eqn:op}), and is given by
$
\Delta_{0} (r) = (8/e^2) \mu (r) \exp[\pi/(2 k_F(r) a_F)],
$
which is valid when $\mu (r) \gg \Delta_{0} (r)$.
Here,
$
\mu (r) = [\mu_\uparrow (r) + \mu_\downarrow (r)]/2
$
is the effective local Fermi energy and
$
k_F (r) = [2M \mu (r)]^{1/2}
$
is the effective local Fermi momentum.
Notice that, $\Delta_0 (r)$ has a maximum at the center of the trap and it
vanishes towards the edges. 
For mass-balanced ($m_\uparrow = m_\downarrow$) mixtures, 
a UPS phase with $Q = 0$ minimizes the free energy when $0 \le |\delta \mu(r)| \lesssim 0.71 \Delta_{0}(r)$,
however, further increase in $|\delta \mu (r)|$ causes a first order transition
to a P$\sigma$PN phase when $|\delta \mu (r)| \gtrsim 0.71 \Delta_{0}(r)$~\cite{clogston, chandrasekhar}.
Therefore, we find in the BCS limit that the UPS fermions are surrounded 
by the P$\sigma$PN fermions, and that they exist only away from the central 
core of the trapping potentials.
For instance, these phases can be seen in Fig.~\ref{fig:trap.di}(b) 
when $1/(k_{F,\downarrow} a_F) = -1$ or $-0.7$, and are schematically shown in Fig.~\ref{fig:trap.shells}(b).
Notice that, similar shell-structures have been also reported for
purely mass-imbalanced mixtures~\cite{duan-mixture, paananen, pao-unequal}.

As the interactions increase, $\Delta_0 (r)$ increases at a faster rate near
$r = 0$ due to the faster increase in local fermion densities, which 
causes an additional first order transition from P$\sigma$PN to UPS at $r = 0$.
For instance, when $\omega_\uparrow = 2 \omega_\downarrow$, this transition occurs 
at $1/(k_{F,\downarrow} a_F) \sim -0.65$ as can be seen in Fig.~\ref{fig:trap.di}(b),
and the shell-structure of this case is schematically shown in Fig.~\ref{fig:trap.shells}(c).
Notice that, this shell-structure does not occur with purely mass-imbalanced 
mixtures~\cite{duan-mixture, paananen, pao-unequal}.
Further increasing the interactions towards unitarity,
we find that the central UPS region expands towards the edges.

In passing to the strongly-attracting limit, we make two comments.
First, it is known that an FFLO-like superfluid phase
with $\Delta_{Q} (r) = \Delta(r) \exp[iQ(r) r]$
and $Q(r) \sim 2.4 M |\delta \mu (r)|/k_F (r)$ may also exist in a small parameter space 
when $0.71 \Delta_{0}(r) \lesssim |\delta \mu(r)| \lesssim 0.75 \Delta_{0}(r)$~\cite{FF, LO}.
This phase resides between the UPS and P$\sigma$PN phases, and is separated 
from the UPS phase by a first order and from the P$\sigma$PN phase by a second order transition.
Notice that, FFLO shells are not shown in Figs.~\ref{fig:trap.shells}(b) and~\ref{fig:trap.shells}(c).
Second, the inclusion of fluctuations beyond the MF 
would reduce $\Delta_0 (r)$, and therefore the transitions discussed above 
are likely to occur at higher values of $1/(k_{F,\downarrow} a_F)$ than our MF predictions. 
While the weakly-attracting MF description is strictly valid 
for $1/(k_{F,\downarrow} a_F) \ll 0$, it still serves as a qualitative estimator for the phase 
boundaries until $1/(k_{F,\downarrow} a_F) \lesssim -0.5$.
However, this description can not be used for $1/(k_{F,\downarrow} a_F) \gtrsim 0$, 
which is discussed next.

{\it Strongly-Attracting Fermion Mixtures}:
In the strong fermion attraction (BEC) limit when $g \gg \{\epsilon_{F,\uparrow}, \epsilon_{F,\downarrow}\}$ or $a_F \gtrsim 0$, 
imbalanced fermion mixtures can be described by a mixture 
of weakly-repulsing Bose molecules and Fermi atoms,
where the Bose molecules correspond to paired $\uparrow$- and $\downarrow$-fermions,
and the Fermi atoms correspond to unpaired fermions~\cite{pieri, iskin-mixture}.
However, in population-balanced ($N_\uparrow = N_\downarrow$) mixtures, 
all $\uparrow$- and $\downarrow$-fermions are paired to form Bose molecules, 
and therefore the equation of motion at $T = 0$ is
\begin{equation}
-\mu_B(r) \Psi_B(r) + U_{BB}|\Psi_B(r)|^2\Psi_B(r) = \frac{\nabla^2 \Psi_B(r)}{2m_B},
\label{eqn:gp}
\end{equation}
which is of the Gross-Pitaevskii form, where 
$
\Psi_B(r) = [M^2 a_F / (8\pi)]^{1/2} \Delta_{0} (r)
$
is the local BEC order parameter,
$
\mu_B(r) = 2\mu(r) -\epsilon_b = \mu_B - V_B(r)
$
is the local chemical potential,
$
U_{BB} = 4\pi a_{BB} / m_B
$
is the repulsive interaction, and
$
m_B = m_\uparrow + m_\downarrow
$
is the mass of the molecular bosons.
Here, 
$
\mu_B = \mu_\uparrow + \mu_\downarrow - \epsilon_b
$
is the chemical potential,
$
V_B(r) = V_\uparrow (r) + V_\downarrow (r)
$
is the trapping potential,
$
\epsilon_b = -1/(M a_F^2)
$
is the binding energy, and $a_{BB} \propto a_F$ is the boson-boson scattering length 
of the molecules. Notice that, identification of
$
V_B (r) = m_B \omega_B^2 r^2/2
$
as the molecular trapping potential leads to
$
\omega_B = [(m_\uparrow \omega_\uparrow^2 + m_\downarrow \omega_\downarrow^2)/(m_\uparrow + m_\downarrow)]^{1/2},
$
which is the effective trapping frequency felt by the molecular bosons.
Therefore, we find in the BEC limit that the ground state of trap-imbalanced 
fermion mixtures is the BEC of molecular bosons for the entire trap.

These results are strictly valid for $1/(k_{F,\downarrow} a_F) \gg 0$, but they still 
serve as a qualitative estimator for the phase boundaries until $1/(k_{F,\downarrow} a_F) \gtrsim 1$. 
For instance, when $\omega_\uparrow = 2 \omega_\downarrow$, we find that the central 
UPS region expands, and the inner UPS and normal regions shrink towards the edges 
with respect to Fig.~\ref{fig:trap.shells}(c), as $1/(k_{F,\downarrow} a_F)$ is increased.
Notice that, the expansion of central UPS towards the edges is similar to the one observed 
with population-imbalanced fermion mixtures~\cite{mit, rice}.
However, in our case, the normal regions vanish beyond a critical $1/(k_{F,\downarrow} a_F)$, 
and the entire mixture becomes a UPS~\cite{pi}.
When $\omega_\uparrow = 2 \omega_\downarrow$, this occurs at $1/(k_{F,\downarrow} a_F) \sim -0.1$,
and the shell-structure of this case is schematically shown in Fig.~\ref{fig:trap.shells}(d).

Having analyzed the ground state phases, next we discuss briefly the validity of 
our results and also their experimental realization in atomic systems.

{\it Experimental Realization}:
In this manuscript, we mainly rely on the LD, MF and isotropic-trap approximations. 
In LD approximation, the mixture is treated as locally homogenous, and this approximation
is valid as long as the number of fermions is large~\cite{torma, liu-mixture, mizushima}, 
which is typically satisfied in atomic systems.
In MF approximation, the superfluid order parameter is treated at the 
saddle-point level, and that the fluctuations are not included~\cite{iskin-mixture}.
This description is qualitatively valid throughout the BCS-BEC evolution
only at low temperatures~\cite{leggett}, which can be reached in atomic systems. 
Lastly, in isotropic-trap approximation, the traps are assumed to be isotropic,
while the atomic traps are typically elongated in one direction.
However, the anisotropy of traps is not expected to affect the shell-structure 
of superfluid and normal phases other than causing shells to have elliptical
rather than circular cross-sections.

In atomic systems, trap-imbalanced fermion mixtures can be realized in several ways. 
For instance, in the case of magnetically trapped systems,
trapping two different hyperfine states ($\uparrow$ and $\downarrow$) of a particular 
atom (i.e., $^6$Li or $^{40}$K) which have different magnetic 
moments (i.e., $\mathcal{M}_\uparrow > \mathcal{M}_\downarrow$) corresponds to a 
situation where $m_\uparrow = m_\downarrow$ and $\omega_\uparrow > \omega_\downarrow$.
Likewise, in optically trapped systems, asymmetrically detuning the laser frequency
with respect to two hyperfine states may produce a state-dependent optical trap.
Furthermore, trap-imbalanced fermion mixtures can be naturally realized with two-species
fermion mixtures~\cite{duan-mixture, paananen, pao-unequal} (i.e., $^6$Li and $^{40}$K) in 
both magnetically and optically trapped systems due to their different mass and also 
to hyperfine properties.

{\it Conclusions}:
We analyzed the ground state phases of two-component population- and mass-balanced 
but trap-imbalanced fermion mixtures as a function of fermion-fermion interactions.
In the BCS limit, we found that the UPS fermions are surrounded by P$\sigma$PN fermions, 
and exist only away from the central core of the trapping potentials.
As the interactions increase towards unitarity, we found that the central P$\sigma$PN core 
first transitions to a UPS, and then expands towards the edges until the 
entire mixture becomes a UPS in the BEC limit.

We thank C. A. R. S{\'a} de Melo, P. S. Julienne, I. B. Spielman and 
R. Grimm for useful discussions.


\begin{thebibliography}{99}
\bibitem{chin} C. Chin et al., Science \textbf{305}, 1128 (2004).
\bibitem{regal2} C. A. Regal et al., Phys. Rev. Lett. \textbf{92}, 040403 (2004).
\bibitem{bourdel1} T. Bourdel et al., Phys. Rev. Lett. \textbf{93}, 050401 (2004).
\bibitem{partridge1} G. B. Partridge et al., Phys. Rev. Lett. \textbf{95}, 020404 (2005).
\bibitem{kinast2} J. Kinast et al., Science \textbf{307}, 1296 (2005).
\bibitem{zwierlein3} M. W. Zwierlein et al., Nature \textbf{435}, 1047 (2005).
\bibitem{leggett} A. J. Leggett, J. Phys. (Paris) \textbf{C7}, 19 (1980).
\bibitem{pao} C. H. Pao et al., Phys. Rev. B \textbf{73}, 132506 (2006).  
\bibitem{sheehy} D. E. Sheehy and L. Radzihovsky, Phys. Rev. Lett. \textbf{96}, 060401 (2006).  
\bibitem{mit} M. W. Zwierlein et al., Science \textbf{311}, 492 (2006).  
\bibitem{rice} G. B. Partridge et al., Science \textbf{311}, 503 (2006).
\bibitem{torma} J. Kinnunen et al., Phys. Rev. Lett. \textbf{96}, 110403 (2006).
\bibitem{pieri} P. Pieri and G. C. Strinati, Phys. Rev. Lett \textbf{96}, 150404 (2006).
\bibitem{yi} W. Yi and L. M. Duan, Phys. Rev. A \textbf{73}, 031604 (2006). 
\bibitem{silva} T. N. De Silva and E. J. Mueller, Phys. Rev. A \textbf{73}, 051602(R) (2006).
\bibitem{haque} M. Haque and H. T. C. Stoof, Phys. Rev. A \textbf{74}, 011602 (2006).
\bibitem{lobo} C. Lobo et al., Phys. Rev. Lett. \textbf{97}, 200403 (2006).
\bibitem{liu-mixture} X.-J. Liu et al., Phys. Rev. A \textbf{75}, 023614 (2007).
\bibitem{mizushima} T. Mizushima et al., J. Phys. Soc. Jpn. \textbf{76}, 104006 (2007).
\bibitem{FF} P. Fulde and R. A. Ferrell, Phys. Rev. \textbf{135}, A550 (1964). 
\bibitem{LO} A. I. Larkin and Y. N. Ovchinnikov, Sov. Phys. JETP \textbf{20}, 762 (1965). 
\bibitem{iskin-mixture} M. Iskin and C. A. R. S{\'a} de Melo, Phys. Rev. Lett. \textbf{97}, 100404 (2006).
\bibitem{pao-mixture} C.-H. Pao et al., Phys. Rev. B \textbf{74}, 224504 (2006).
\bibitem{duan-mixture} G.-D. Lin et al., Phys. Rev. A \textbf{74}, 031604(R) (2006).
\bibitem{parish} M. M. Parish et al., Phys. Rev. Lett. \textbf{98}, 160402 (2007).
\bibitem{paananen} T. Paananen et al., Phys. Rev. A \textbf{75}, 023622 (2007).
\bibitem{pao-unequal} C.-H. Pao et al., cond-mat/07083167.
\bibitem{clogston} A. M. Clogston, Phys. Rev. Lett. \textbf{9}, 266 (1962).
\bibitem{chandrasekhar} B. S. Chandrasekhar, Appl. Phys. Lett. \textbf{1}, 7 (1962).
\bibitem{pi} In the strongly-attracting limit of a population- and trap-imbalanced mixture, 
unpaired excess fermions (paired bosons) are pushed away from the center of the trapping 
potentials when fermions with loose (tight) trap is sufficiently in excess.

\end{thebibliography}
\end{document}